\documentclass[onecolumn]{rsauthor}

\def\msun{M_\odot}
\def\msunyr{M_\odot \ {\rm yr}^{-1}}

\def\la{\mathrel{\mathpalette\fun <}}
\def\ga{\mathrel{\mathpalette\fun >}}

\def\simpropto{\lower.2ex\hbox{$\; \buildrel \sim \over \propto \;$}}

\def\fun#1#2{\lower0.837ex\vbox{\baselineskip0ex\lineskip0.209ex
  \ialign{$\mathsurround=0ex#1\hfil##\hfil$\crcr#2\crcr\sim\crcr}}}

\jname{Phil. Trans. R. Soc. A}

\begin{document}

\title{High Energy Transients}

\author{Neil Gehrels$^{1}$, John K. Cannizzo$^{1,2}$} 

\address{$^1$Astroparticle Physics Division, NASA/Goddard Space
       Flight Center, Greenbelt, MD 20771, USA \\
$^2$CRESST/Joint Center for Astrophysics, Univ. of Maryland, 
    Baltimore County, Baltimore, MD 21250, USA }

\abstract{We present an overview of
    high energy transients in astrophysics,  
    highlighting important advances over the past
 50 years.
    We begin with early discoveries of
  $\gamma-$ray transients,
   and then delve into physical details
  associated with a variety of phenomena.
   We discuss some of the unexpected transients
  found by {\it Fermi} and {\it Swift},
  many of which are not easily  classifiable
  or in some way challenge conventional wisdom.
These objects are important insofar as they underscore 
the necessity of future, more detailed studies.
  }


\date{Received xx Month1 2012; Accepted yy Month2 2012}

\maketitle

\section{Early Discoveries in $\gamma-$ray Astronomy: GRBs, Solar Flares,
  and Supernovae}\label{sec1}

Some of the early 
   discoveries in
   $\gamma-$ray 
   astronomy 
         were unexpected
 for different reasons.
   Some unveiled
     entirely new phenomena,
  whereas
  others showed that $\gamma-$rays
   could 
     accompany  previously known phenomena
   that had not been  suspected to have a high energy component.
The discovery of $\gamma-$ray bursts  (GRBs) in the 1960's
was serendipitous, and falls into the first category.
     The \emph{Vela} satellites were launched to verify 
the 1963 Partial Test Ban Treaty governing the testing of
nuclear weapons. They contained  $\gamma-$ray and X-ray detectors.
On July 2, 1967 a flash of $\gamma-$radiation unlike that expected
from nuclear testing was observed by a team led by Ray Klebesadel.
  Years later when the 
  puzzling results were fully analyzed and understood  
 they became the basis of the discovery paper for GRBs 
  (Klebesadel et al. 1973).
  For many years the distance scale to GRBs was unknown, prompting
theories ranging from local (i.e., solar system) to 
extragalactic.  The  next breakthrough came
  with the 
   \emph{Compton Gamma Ray Observatory}
   (\emph{CGRO}) which 
   discovered  more than 2600 bursts in just 9 years (1991-2000)
 and provided localizations of $\sim$$3^{\circ} - 20 ^{\circ}$
  for individual bursts.
    Their isotropy over the sky hinted at a cosmological origin.
In 1997 the first  $\sim$1 arcmin
     localizations
   (done by \emph{BeppoSAX})
  led to the identifications of galaxies within which GRBs had occurred,
 and subsequent
     redshift determinations showed them to lie at cosmological distances
  (van Paradijs et al. 1997; Bloom et al. 2001).

The discovery of $\gamma-$rays in solar flares
      falls into the latter category: energetic radiation
from an unexpected source.   
A solar flare is an explosion in the solar atmosphere 
  due to the sudden release of magnetic energy 
 in the corona. Flares occur
in active regions around sunspots  where
  strong magnetic field lines penetrate the
  photosphere and connect the corona to the solar
interior.
 Solar flares can be quite energetic,
  up to $\sim0.16 L_\odot$,
 releasing energy  
  across the full electromagnetic spectrum 
from the radio waves
 to high energy $\gamma-$rays. 
The first detection of 
$\gamma-$radiation from a
  solar flare was August 4, 1972 by
{\it OSO-7} using a
 3"$\times$3" NaI crystal
  detector
  (Chupp et al. 1973).

\begin{figure}[h!]  
\begin{centering}
 \includegraphics[height=4.25in]{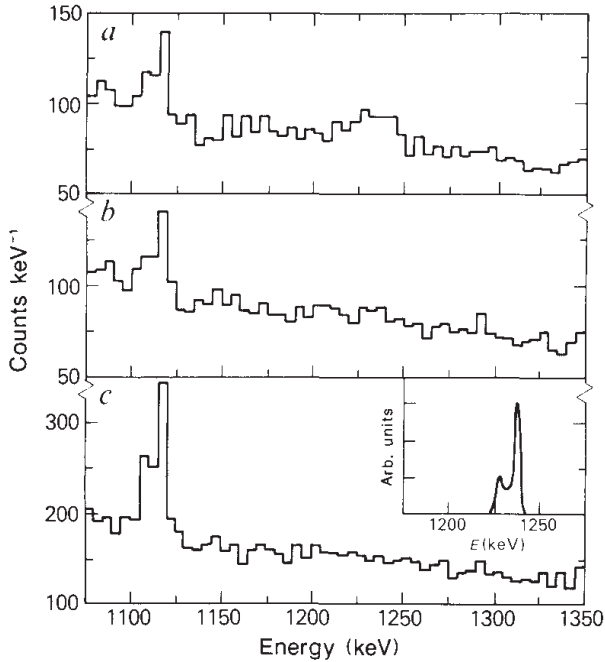}
 \caption{GRIS spectra of 1987A (Teegarden et al. 1989).}
\end{centering}
\end{figure}

  The discovery of $\gamma-$rays
  in supernovae (SNe) was not unexpected,
   but the early arrival after the initial SN blast
was.
  High energy photons had been predicted 
as a result of the shock breakout from the
  stellar interior. 
The discovery of SN 1987A in the LMC on Feb 24, 1987
gave astronomers a ringside seat to the nearest
SN in 400 yrs.
     The
     detection of a  $\gamma-$ray signal in
         SN 1987A by \emph{SMM} (Matz et al. 1988)
 and the GRIS balloon flight
 from Alice Springs, Australia on 1 May 1988
   (Teegarden et al. 1989) 
  at  a little over one year after the SN,
      was much sooner than expected
  and prompted theorists to propose that  
  ``fingers'' of ejecta from the exploding core
        are able to penetrate
  the  overlying, expanding stellar layers
  more rapidly that the earlier, spherically
  symmetric estimates had indicated\footnote{The
  detection of a hard X-ray continuum
  in 
   Aug 1987
  (Dotani et al. 1987, Sunyaev et al. 1987)
 led to a revision
  in the theoretical prediction
   for the time of appearance of the $\gamma-$ray signal
   (Pinto \& Woosley 1988ab).}.
    The GRIS observations 
 were made on day 433 after the SN, whereas a spherically
  symmetric shock breakout was not expected by theorists
    for several years.
  Teegarden et al. observed line emission at 1.238 MeV
  with a full width at half maximum of $16.3\pm6$ keV (Figure 1).
 The optical light curve of SN1987A indicated a production
of $0.075\msun$ of radioactive $^{56}$Ni  in the initial
  explosion.
   Comparison of the radioactive output of the daughter
product $^{56}$Co  with the measured GRIS  1.238 MeV
  line flux indicates only $\sim$13\% of the  1.238 MeV $\gamma-$rays
  escaping $-$ under the assumption of spherical symmetry.

\section{The Modern Era}\label{sec2}

\subsection{Missions}

Currently operating satellites
carrying $\gamma-$ray detectors
 benefit from decades of experience 
  plus technological advances.
  Three of the mainstays in  present day
 $\gamma-$ray astronomy are
  {\it INTEGRAL}, {\it Swift}, and {\it Fermi}.

 \emph{INTEGRAL}
({\it INTErnational Gamma-Ray Astrophysics
   Laboratory} $-$ Winkler et al. 2003)
    was launched in 2002
into a 72 hr orbit with a perigee of 
  10,000 km and an apogee of 153,000 km.
  It was four coaligned instruments:
(1) The IBIS (Imager on-Board the \emph{INTEGRAL} Satellite)
   observes between $\sim$15 keV and $\sim$10 MeV,
and an angular resolution of $\sim$12 arcmin.
It consists of a 95$\times$95 mass of rectangular
 tungsten tiles 3.2 m above a detector consisting of
128$\times$128 CaTe tiles backed by 64$\times$64 CsI tiles.
  The detectors are surround by tungsten/lead shielding.
(2) The main spectrometer, the SPI (SPectrometer for
 \emph{INTEGRAL}) is sensitive from $\sim$20 keV to $\sim$8 MeV
  and  comprises a coded mask of hexagonal
 tungsten tiles overlying a detector plane of 19 Ge
  crystals. The resolution is $\sim$2 keV at 1 MeV.
(3) The ACS (AntiCoincidence Shield) is composed
  of  a mask shield of plastic scintillator
    behind a detector shield of tungsten tiles and BGO scintillator
 tiles. The all-sky coverage of the ACS make it a valuable
GRB detector, and one of the components of the IPN
   (InterPlanetary Network) for localizing GRBs.
(4) There are dual JEM-X units that observe from 3 to 35 keV
   using gas scintillation detectors in a microstrip layout.

  {\it Swift} (Gehrels et al. 2004)
    was launched in 2004 into a standard low Earth orbit 
   (600 km). It carries three science instruments:
(1)  The BAT (Burst Alert Telescope) is a coded-aperture
    mask of 52,000 randomly placed 5 mm Pb tiles situated
  1 m above a detector plane of 32,768 4 mm CdZnTe tiles. 
  It covers $>1$ sr fully-coded ($\sim$3 sr partially coded)
 and detects and localizes GRBs to $\sim$$1-4$ arcmin
 within 15 s.  Its energy range is $15-150$ keV.
(2)  The XRT (X-ray Telescope) is one of two
  co-pointed NFI (narrow field
instruments) that are trained on a GRB after the spacecraft
has slewed to the BAT determined localization.
   The XRT uses a Wolter Type I X-ray telescope with 12 nested
mirrors focused onto a single MOS  CCD.
   It has sensitivity in the range $0.2-10$ keV.
(3) the UVOT (Ultraviolet/Optical Telescope) is
   the other NFI which is used to study GRB afterglow.
     It can obtain positions to less than one arcsec,
  and provides optical and UV photometry and
  low resolution spectra in the range $170-650$ nm.

{\it Fermi} (formerly the \emph{Gamma-ray Large Area Space Telescope} $-$
 \emph{GLAST}
   $-$ Atwood et al. 2009) was launched 2008 into a standard low Earth orbit
  (550 km). It has two instruments:
(1) The LAT (Large Area Telescope), an imaging $\gamma-$ray
detector has a FoV (field of view) of $\sim$2.4 sr
and sensitivity between $\sim$30 MeV and $\sim$300 GeV.
   It is a natural successor to the EGRET instrument
  on the {\it Compton Gamma-Ray Observatory}.
(2)  The GBM (Gamma-ray Burst Monitor) can detect GRBs over
the entire non-Earth occulted sky and has sensitivity from 
$\sim$8 keV to $\sim$30 MeV. It consists
of 14 scintillation detectors $-$ 12 NaI crystals with energy
range $\sim$8 keV $-$ $\sim$1 MeV and two  BGO crystals
with energy range $\sim$150 keV $-$ $\sim$30 MeV.

\begin{table}[h!]
\centering
\small
\caption{High Energy Transients} 
\begin{tabular}{@{}cccc@{}}
\hline
 Source       &  typical duration     & Energy Source & E($\gamma-$ray)   \\
\hline
TGF            & msec                  & E field     &   $10^{10}$ erg          \\
GRB            & msec $-$ mins          & gravity     &  $10^{51}$ erg          \\
SGR            & msec $-$ sec            & B field     & $10^{44}$ erg          \\
TDE            & day - yrs                & gravity    & $10^{52}$ erg          \\
solar flare     & mins                     & B field   & $10^{32}$ erg          \\
SN/nova          & mins $-$ yrs             & nuclear  & $10^{49}$ erg          \\
accreting BH/NS  & secs $-$ days (variable)  & gravity & $10^{36}$ erg s$^{-1}$ \\
AGN              & hrs   $-$ days (variable) & gravity & $10^{43}$ erg s$^{-1}$ \\
\hline
\end{tabular} \label{ta1}
\end{table}

\subsection{Science}

Table 1 indicates the panoply of high energy transients,
  with their associated timescales and energies.

\begin{figure}  
\begin{centering}
 \includegraphics[height=4.05in]{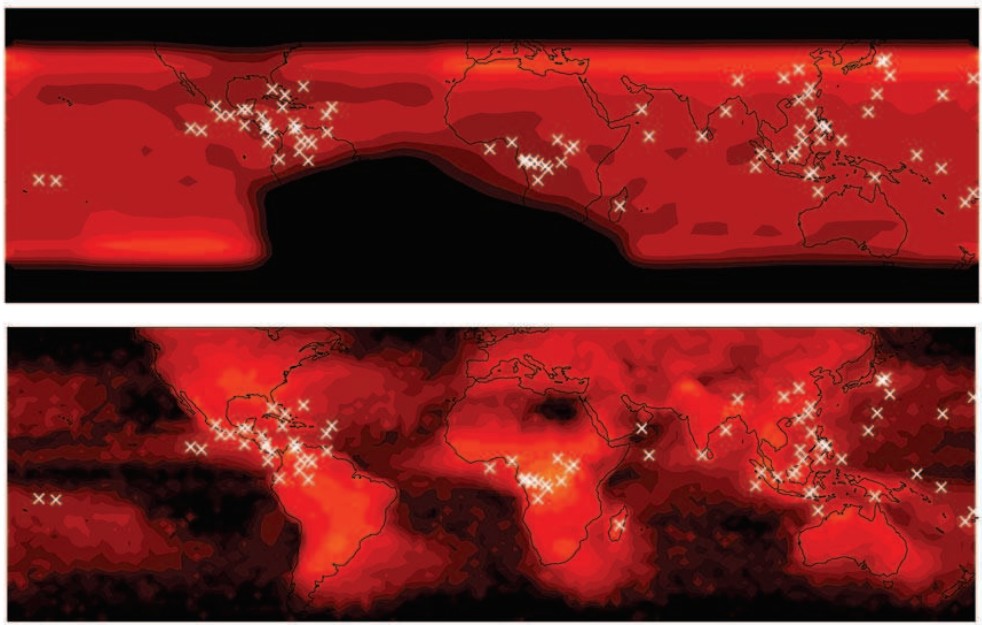}
 \caption{\emph{RHESSI} position projected onto the Earth's
  surface during each recorded TGF, plotted over 
  (i) the expected distribution of observed TGFs 
  if the population were evenly distributed
over the globe,
  and (ii) the long-term lightning frequency data (Smith et al. 2005).} 
\end{centering}
\end{figure}

 Terrestrial $\gamma-$ray flashes (TGFs) were first
seen the the Burst and Transient Source Experiment (BATSE) 
  on the {\it Compton Gamma-Ray Observatory}.
 They are thought to be due to decaying electric fields
above  thunderclouds after a lightning discharge.
  Relativistic electrons interact with nuclei 
of atoms in the atmosphere and produce $\gamma-$rays
via bremsstrahlung. A process known as 
   runaway electron avalanche is thought to be relevant, 
but the details are uncertain (Gurevich et al. 1992, Dwyer 2003). 
 Subsequent observations by the \emph{Reuven Ramaty High Energy Solar 
 Spectroscopic Imager} (\emph{RHESSI})
  have revealed TGFs with much higher energies,
   and show that $\sim$500 TGFs occur per day (Figure 2, from Smith et al. 2005),
   which is a small fraction of the total number of daily
  lightning strikes on Earth ($\sim$$3-4\times 10^6$).
     This estimate excludes effects relating to beaming and
   atmospheric obscuration of low altitude TGFs.
   The \emph{Fermi}/GBM is currently detecting
TGFs at a rate of $\sim$10  yr$^{-1}$; 
       some are even detected by \emph{Fermi}/LAT during special
Earth-pointed observations.

Gamma-ray bursts (GRBs) are intense flashes of radiation
 produced at cosmological distances $z\simeq 2$.
 GRBs come in two primary flavors, long and short,
with the dividing point being roughly 2 s (Kouveliotou et al. 1993).
 A further division
can be made spectrally according to their hardness ratio
(i.e., ratio of high to low energies).
The redshift range is from
  about 0.2 to 2 for short GRBs (sGRBs),
  with a mean of about 0.4.
  For
  long GRBs (lGRBs)
  the range is between about 0.009 and 8.2,
  with a mean of about 2.3.
The typical energy release is $\sim$10$^{49}-10^{50}$ erg
 for sGRBs and  $\sim$10$^{50}-10^{51}$ erg for lGRBs.
  These ranges are based on observed isotropic-equivalent
  energies of  $\sim$10$^{51}$ erg for sGRBs and
               $\sim$10$^{53}$ erg for lGRBs,
   and
   estimates for jet beaming
for each class,
      $\theta_{\rm j}\sim5^{\circ}$ for lGRBs
          and
     $\theta_{\rm j}\sim5-15^{\circ}$ for sGRBs
 (Burrows et al. 2006, Grupe et al. 2006, Fong et al. 2012).
  Beaming angles for sGRBs
  are still highly uncertain.
  The
  corresponding
  beaming factors
 $f_b = 1-\cos\theta_{\rm j}\simeq \theta_{\rm j}^2/2$
 are  roughly $1/300$ for lGRBs and $1/30$ for sGRBs.
 The $L_{\rm X}/E_{\gamma-{\rm iso}}$ values 
  at 11 hr post-GRB 
   are similar between lGRBs and sGRBs (Gehrels et al. 2008).
 The sGRBs have weaker X-ray afterglows,
   a mean value of
 $\sim$7$\times 10^{-10}$ erg cm$^{-2}$ s$^{-1}$  versus
 $\sim$3$\times 10^{-9}$  erg cm$^{-2}$ s$^{-1}$ for lGRBs.
Although  many of the details 
  are uncertain, the two 
  mechanisms  are thought to be collapsars for
  long GRBs (lGRBs) and merging neutron stars for short GRBs (sGRBs)
 (Gehrels, Ramirez-Ruiz, \& Fox 2009).
  LGRBs are intrinsically very bright,
the brightest explosions in the Universe.
  The highest redshift lGRBs were far above detector
  threshold.
 SGRBs are intrinsically less luminous by a factor $\sim$$10^3$,
and are seen primarily nearby, $z\la1$.

Soft gamma repeaters (SGRs) emit
bursts and $\gamma-$rays and X-rays
which are thought to be
due to the rearrangement of
powerful magnetic fields
in magnetars $-$ 
pulsars with magnetic fields of 
$\sim$$10^{15}$ G.
  The first one seen was the ``March 5th Event'' from 1979
  which was
observed by two Soviet
interplanetary
  spacecraft
  {\it Venera 11} and  {\it Venera 12}
 (Mazets et al. 1979).
   This event, SGR 0526-66, was localized to 
the SN remnant N49
in the LMC. Given a distance of $\sim$50 kpc, the isotropic
equivalent energy emitted was $\sim$$5\times 10^{44}$ erg,
compared to $\sim$$10^{41}$ erg for a typical SGR burst.
   A $\sim$8 s periodicity thought to be the NS spin period
is plainly evident in the data.
  Thompson \& Duncan (1995) present an extensive
  model of the March 5th event and
other SGRs as magnetars, i.e., $|B|\simeq 10^{15}$ G.
   They argue that the March 5 burst was due to a
large-scale readjustment of the stellar magnetic
field,
  while the more standard SGR bursts 
         are caused by the release
of magnetic stresses within a more localized 
      patch of the crust.
    Thompson \& Duncan put forth a variety of
independent arguments in favor of the magnetar scenario,
including (i) the necessity of a very high $B-$field to
spin down the pulsar to $\sim$8 s within the inferred
$\sim$$10^4$ yr age of N49, (ii) a very strong $B-$field
   suppresses the $e^-$$-$scattering cross section
below the standard Thomson value
by the ratio $\sim(B/B_{\rm QED})^{-2}$,
  where $B_{\rm QED} = m_e^2 c^2/(e\hbar)=4.4\times 10^{13}$ G
(the point at which the nonrelativistic Landau energy
   $\hbar eB/(m_e c)$ equals 
      the electron rest energy $m_e c^2$),
therefore enabling $L\simeq 10^4L_{\rm Edd}$
for surface fields $\ga10^{14.5}$ G,
  (iii) persistent X-ray emission from SGR 0526-66
at $\simeq 7\times 10^{35}$ erg s$^{-1}$ (Rothschild 
  et al. 1993, 1994) implies $B_{\rm crust} \ga 10^{15}$ G,
   and
  (iv) an identification
  of the $\sim0.15$ s duration of the hard spike 
 of the March 5 event
 (Mazets et al. 1979,
   Cline et al. 1980)
      with the internal Alfv\'en
 crossing time leads to $B\simeq 7\times 10^{14}$ G.

\begin{figure}[h!]  
\begin{centering}
 \includegraphics[width=5.25in]{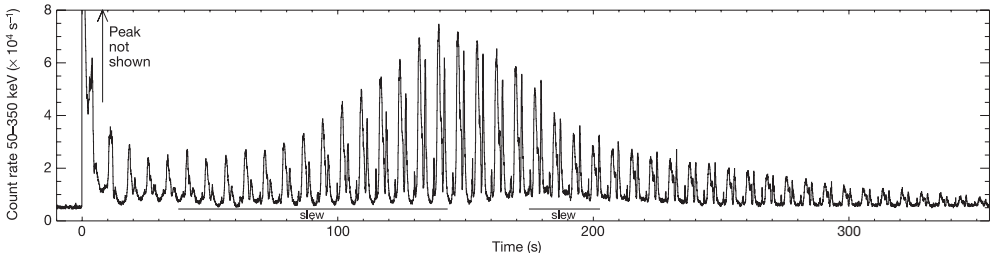}
 \vskip -1truein
 \caption{The spike and tail light curve
  for SGR 1806-20 from \emph{Swift}/BAT (Palmer et al. 2005).}
   \label{figxxx}
\end{centering}
\end{figure}

On Aug 27, 1998 a second giant flare from
  an SGR was seen.  SGR 1900+14
became the brightest
extra-solar system
 $\gamma-$ray source
  ever.  The 5.16 s spin period of
the pular could be easily seen
  directly
 in the light curve (Kouveliotou et al. 1999), and produced
ionization changes in the upper atmosphere
of the Earth.
  Thompson \& Duncan (2001)
 argue that the extremely high luminosity 
  $L\ga 10^6L_{\rm Edd}$ during the initial
  $\sim$0.5 s spike in SGR 1900+14 demands  $B\ga10^{15}$ G.
  Shortly after the launch of {\it Swift}, on 27 Dec 2004,
  a giant $\gamma-$ray flare was seen from SGR 1806-20
  (Figure 3, from Palmer et al. 2005) with a peak flux of 
  $\sim$5 erg cm$^{-2}$ s$^{-1}$.
 SGR bursts
  are much brighter than ordinary X-ray
bursts, which are due to thermonuclear flashes
of accumulated
hydrogen on the surface  of a NS 
($L\sim10^3-10^4$$L_{\rm Edd}$ vs. $L\sim$$L_{\rm Edd}$)
and have harder spectra.

Tidal disruption events (TDEs)
   are caused by the tidal disruption of
stars that venture too close to the massive black holes (MBHs)
at the centers of galaxies (Rees 1988, Phinney 1989).
   Prior to March 2011, nearly all our observational
   information was based on optical/UV studies 
  (Gezari et al. 2006, 2008)
    or 
  long-term X-ray data with poor time sampling (Komossa et al. 2004).
  This changed with the discovery  by {\it Swift}
 of GRB 110328A/Swift J1644,
  a TDE viewed down the jet axis of a MBH 
in the nucleus of a galaxy
  at redshift $z=0.35$ (Bloom et al. 2011,
  Burrows et al. 2011, Levan et al. 2011).
 Continued observations  for over 1 yr with the 
{\it Swift}/XRT has shown an apparent long term
decay law $L_x \propto t^{-\alpha}$
with $\alpha \simeq -1.3$,  
   which may be consistent with the decay of
 a freely expanding, advectively dominated
  slim disk (Cannizzo, Troja, \& Lodato 2011).
  The this decay law appears to hold as early
as $t\simeq 10$ d, indicating that the conventional
 dividing point between
 ``stellar     fallback'' ($L\propto t^{-5/3}$)
   and ``disk accretion'' ($L\propto t^{-4/3}$)
  (Phinney 1989, Cannizzo, Lee, \& Goodman 1990)
   may have been at $\la 10$ d,
   indicative of a deeply plunging disruption.
   This is in contrast to the more probable event
   where a disruption occurs
   close to the classical tidal disruption radius,
   in which case the dividing point would lie at  years to decades.
 If Sw1644 was deeply plunging, that may also be part of the reason
  it was a powerful, jetted TDE.

\begin{figure}[h!] 
\begin{centering}
\includegraphics[height=3.05in]{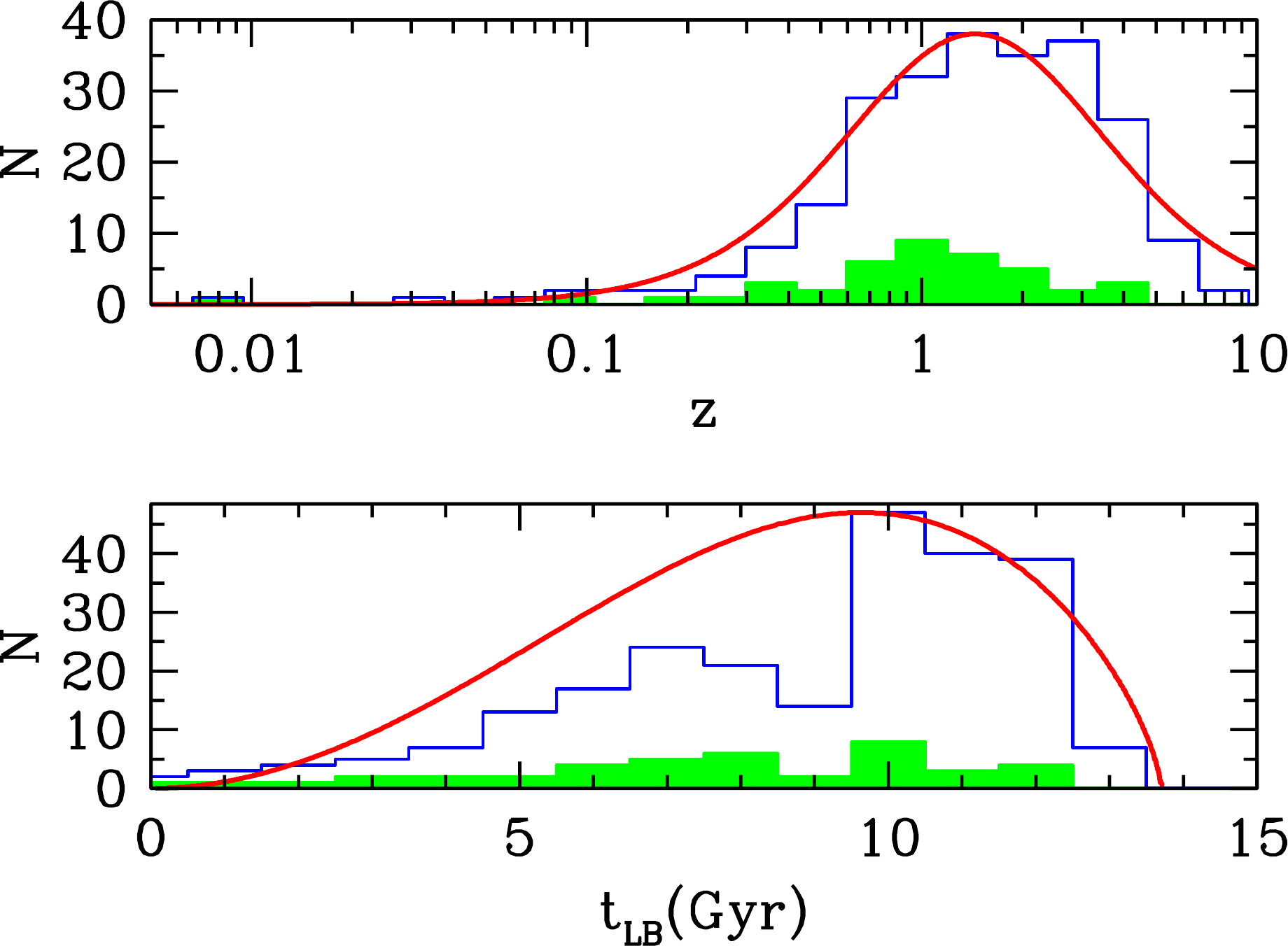}
\caption{The frequency histogram distribution
     for all spectroscopic GRB redshifts
     determined to date (243 - shown in blue)
     as well as the pre-{\it Swift} era
 distribution   (41 - shown in green).
    The top panel shows the distribution
   in $z$, and the bottom panel in  look-back time.
   The red curves indicate the evolution of
 a slice of comoving volume $(dV/dz)(1+z)^{-1}$,
 where the factor $(1+z)^{-1}$
   corrects for cosmological time dilation,
   given that the GRB rate has effective
units of volume$^{-1}$ time$^{-1}$.}
\end{centering}
 \end{figure}

\section{GRBs in the {\it Swift} Era}

{\it Swift}
   Has brought brought about a revolution
 in GRB research. Redshifts from GRBs
   discovered prior to {\it Swift} amount to 41 total. 
   (At the time of {\it Swift}'s launch this number was
 $\sim$25, but continued observation of identified host galaxies
with time has increased the pre-{\it Swift} total.)
   Now there are over 200 redshifts.
  The total number of  {\it Swift} GRBs is approaching 700.
   Figure 4 shows a plot of frequency histogram distribution
   for redshifts,
   excluding uncertain values and photometric redshifts. 
 As a first approximation, the GRB rate history traces
    a the volume of the universe.
 Of the 243 total redshifts we have now, 187 are from GRBs
   discovered by {\it Swift}.
There are 18 from {\it HETE}, 15 from {\it BeppoSax}, 
  10 from {\it IPN},
   6 from {\it Fermi},
and 4 from {\it INTEGRAL}.

\section{Oddball Events}

{\bf Short ``GRB'' 050925:}
  This unusual burst triggered the BAT with a single peaked
outburst of duration $T_{90}=70$ ms 
  (Holland et al. 2005, Markwardt et al. 2005).
 It occurred near the galactic plane, and nothing was seen in the UVOT.
   However, the $V-$band extinction toward the source was $A_V = 7.05$ mag.
 The XRT spectra and lightcurve show no significant X-ray emission
in the field, suggesting that any X-ray counterpart to this burst was faint.
  Markwardt et al. (2005) were able to fit a power law spectrum to the BAT
data, with a photon index $1.74 \pm 0.17$;
  they found a better fit could be obtained with a blackbody spectrum
  $kT = 15.4 \pm 1.5$ keV,
   a value consistent with small-flare events from SGRs.
  The low galactic latitude and soft spectrum indicate a possible
 galactic source or SGR.

{\bf GRB 060218/SN 2006aj:}
On 18 February 2006 \emph{Swift} detected the remarkable burst 
 GRB 060218 that provided considerable new information on the 
 connection between SNe and GRBs.  It was longer (35 min) and 
 softer than any previous burst, 
   and was associated with 
 SN 2006aj at only $z=0.033$.
   SN 2006aj was  a (core-collapse) SN Ib/c
with an isotropic energy equivalent of
a few $10^{49}$ erg,
   thus underluminous compared
   to the overall energy
   distribution for long GRBs.
The spectral peak in prompt emission at $\sim$ 5 keV places 
 GRB 060218 in the X-ray flash category of GRBs (Campana et al. 2006), 
 the first such association for a GRB-SN event.  Combined BAT-XRT-UVOT 
 observations provided the first direct observation of shock-breakout 
 in a SN (Campana et al. 2006). 
   This is inferred from the evolution of a 
 soft thermal component in the X-ray and UV spectra, and early-time 
 luminosity variations.  Concerning the SN,
   SN 2006aj was dimmer 
by a factor $\sim$2 than the previous SNe associated with GRBs, 
 but still $\sim$$2-3$ times brighter than normal SN Ic not
   associated with GRBs 
  (Pian et al 2006, Mazzali et al 2006).
GRB 
    060218 was an underluminous burst, as were two of the other 
 three previous cases.  Because of the low luminosity, these events are 
 only detected when nearby and are therefore rare.  
  However, they are actually  $\sim$$5-10$ 
 times more common in the universe 
  than normal GRBs (Soderberg et al. 2006).

\begin{figure}[h!]   
\begin{centering}
 \includegraphics[height=2.85in]{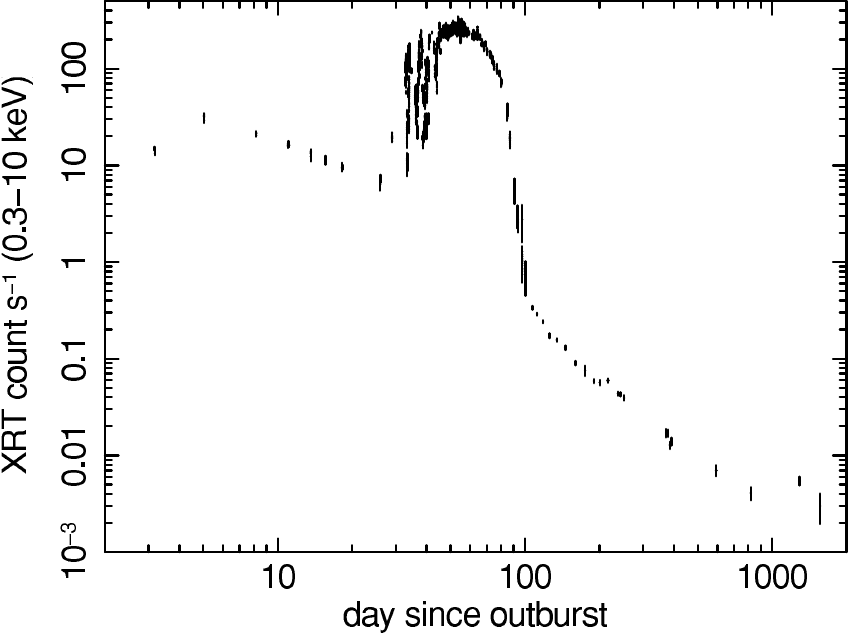}
 \caption{The 2006 nova
   outburst from RS Oph (Osborne et al. 2011).}
   \label{fig3}
\end{centering}
\end{figure}

{\bf RS Oph 2006:}
Novae can occur in interacting binaries
  containing a WD accretor
 and are due to
the thermonuclear detonation of
    accreted material on the
   surface of a WD
  (Gallagher \& Starrfield 1978).
  This can occur if the temperature
  and pressure at the base of
  the accumulated layer
  of accreted matter
    are in the appropriate regime.
    \emph{Swift} has opened a new
   window on nova studies.
   To date  \emph{Swift} has observed 28 novae.
   It has detected keV emission from shocked
ejecta and supersoft (SS) emission from the WD surface.
  Extensive observations ($\sim$400 ks)
   of the 2006 nova outburst
  from RS Oph  (Figure 5)
   found an unexpected
  SS state,
     and 35 s QPO (Osborne et al. 2011).
    Detailed analysis of \emph{Swift}
  observations  revealed a mass
ejection of $\sim$$3\times10^{-5}M_\odot$
  at $\sim$$4000$ km s$^{-1}$
   into the wind of the mass losing red
  giant companion in the system.

{\bf GRBs 060505 \& 060614: }
GRB 060505 and 
 GRB 060614 were  nearby  GRBs ($z=0.089$ and 0.125, resp.)
   with no coincident SN, to deep
limits (Fynbo et al. 2006).
  GRB 060614 was bright ($15-150$ keV fluence of $2.2\times 10^{-5}$
erg cm$^{-2}$), and, with a $T_{90}$ of 102s, seemed to be a
secure  long GRB.   Host galaxies were found (Gal-Yam et al. 2006,
    Fynbo et al. 2006, della Valle et al. 2006)
  and deep searches were made for  coincident SNe.
  All other well-observed nearby GRBs have had SNe, 
  but GRB 060614 did not to limits $>100$ times fainter
  than previous detections (Gal-Yam et al. 2006, Fynbo et al. 2006,
  della Valle et al. 2006).
 GRB 060614 shares some characteristics with sGRBs (Gehrels et al. 2006).
    The BAT light curve shows an initial short hard flare
lasting $\sim$5 s, followed by an extended softer episode, $\sim$100 s.
   The light curve is similar to some \emph{Swift} sGRBs and
a subclass of BATSE sGRBs (Norris \& Bonnell 2006).
 GRB 060614 also falls in the same region
of the lag-luminosity diagram as sGRBs (Figure 6).   
 Thus GRB 060614 is problematic to classify.
      It is a lGRB by the traditional definition,
  but lacks an associated SN. It shares some similarities
with sGRBs, but the soft episode is brighter, which would be
difficult to account for in the NS-NS merger scenario.

\begin{figure}[h!]
\includegraphics[width=5.25in]{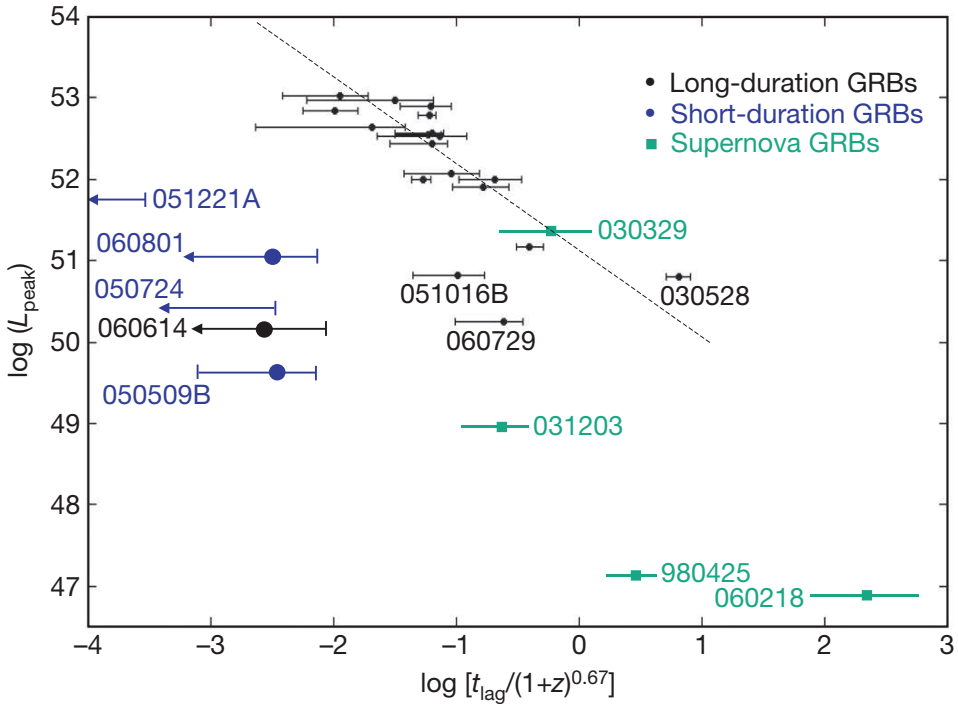}
\caption{Spectral lag as a function of peak luminosity showing
GRB 060614 in the region of short-duration GRBs
  (Gehrels et al. 2006).}
 \end{figure}

{\bf Hostless GRB 070125:}
  There was not an obvious
  host galaxy for  GRB 070125. 
   Deep ground-based
 imaging reveals no host 
 to  $R > 25.4$ mag.
 Cenko et al. (2008)
 present an analysis of spectroscopic data
which reveals weak Mg II lines indicative of halo gas.
In the field are two blue galaxies offset by $\ga$27 kpc
at $z=1.55$. If there is an association with one of them, 
it would imply a velocity $\sim$$10^4$ km s$^{-1}$
  over a $\sim$20 Myr lifetime of the massive progenitor.
  The only known way of achieving this would have been
a prior close interaction with a massive BH.
  However, this interpretation
was muddied by Chandra et al. (2008),
 who inferred a dense environment, based
on bright, self-absorbed radio afterglow.
They proposed a scenario in which the high
density material lies close to the explosion site,
 and the lower density material further away.
 They note  GRB 070125 was one of the brightest
GRBs ever detected, with an isotropic
 release of $10^{54}$ erg
(by comparison, $\msun c^2\simeq 2\times10^{54}$ erg).
The prompt emission from GRB 070125
was also seen by \emph{Suzaku}/WAM (Onda et al. 2010).

\begin{figure}[h!] 
\includegraphics[width=5.25in]{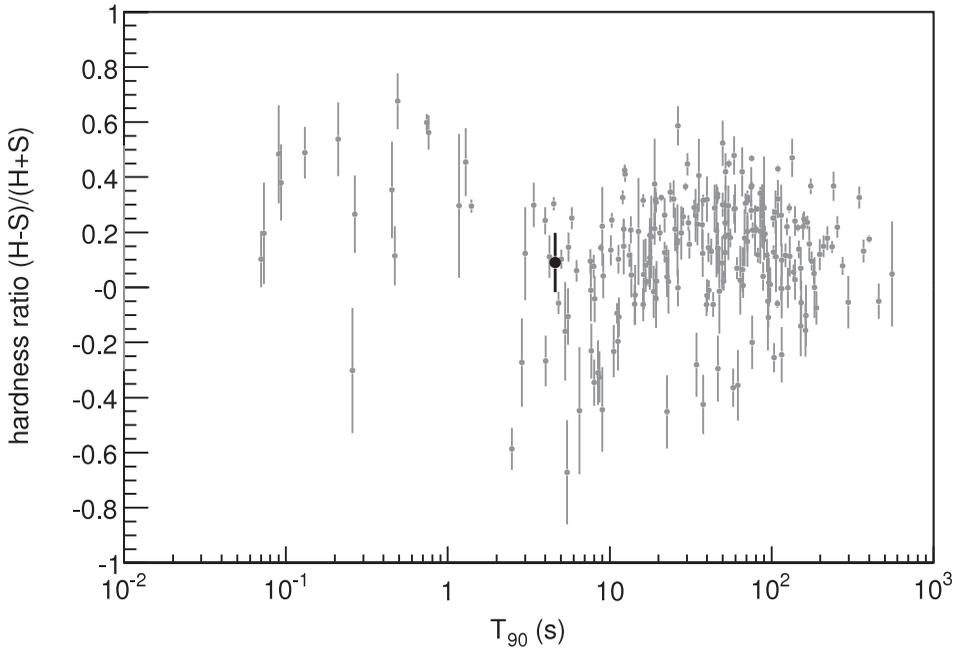}
\caption{Duration ($T_{90}$) and hardness ratio ($HR$) of 226 Swift bursts from
GRB 041217 to GRB 070616, where $HR\equiv (H-S)/(H+S)$,
  and
 $S$ and $H$ are energy fluences in the $15-50$ and $50-150$ keV
  bands, respectively
  (Kasliwal et al. 2008). The
values of $T_{90}$ and $HR$ GRB 070610 
  (large filled black circle) are
$4.6\pm0.4$ s and $0.09 \pm 0.11$, respectively.}
 \end{figure}

\begin{figure}[h!]   
\begin{centering}
\includegraphics[height=3.35in]{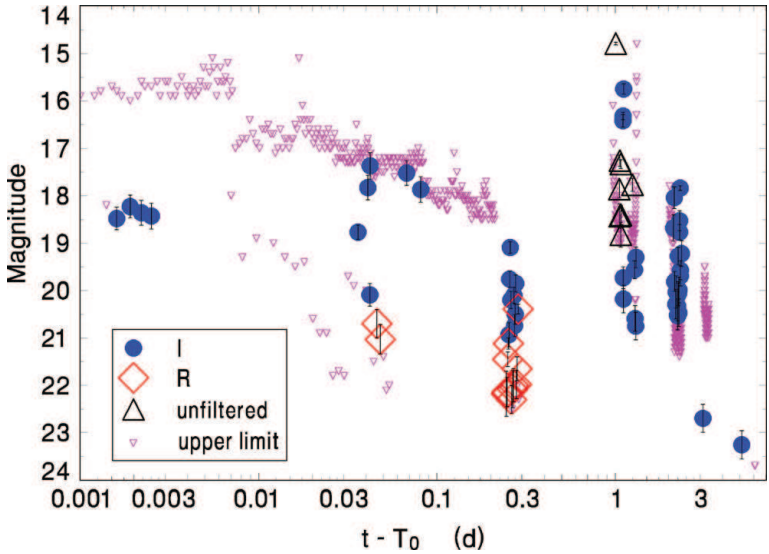}
\caption{The light curve of SWIFT J1955+2614
  in several photometric bands for
 $t -T0
 < 7$ d, where $T0$ is the time of GRB 070610
  (\v{S}imon et al. 2012).} 
\end{centering}
 \end{figure}

{\bf Galactic ``GRB'' 070610:}
 Discovered initially as GRB 070610,
  this object, now dubbed
  Swift J195509.6+261406 (Sw1955+26),
  is thought to represent a member of a
relatively new class of fast X-ray nova 
 containing a BH.  It had a duration of $\sim$5 s,
 and also shows large variability.
  Kasliwal et al. (2008) 
   discuss several possibilities for
this source (Figure 7), and propose an analogy
with V4641 Sgr,
   an unusual BH binary
which had a major outburst in 1999
  (in't Zand et al. 2000).
 V4641 Sgr is a binary with a B9 III star
orbiting a $\sim9$$\msun$ BH
  (Orosz et al. 2001)
  and also exhibited strong and fast
X-ray and optical variability.
  The analog is imperfect in that the normal star
 in  Sw1955+26 is a cool dwarf rather than a B9 giant,
  suggesting a physical origin for the bursting
 behavior in the accretion disk and/or jet
rather than the mass donor star.
  A $\sim$5 s burst is certainly distinct from
the $\sim$month long fast-rise exponential decay
 seen in systems like A0620-00 in 1975
  (Tanaka \& Shibazaki 1996,
  Remillard \& McClintock 2006)
  which are thought to be due to a large
scale storage and dumping of material
 in an accretion disk 
   (Cannizzo et al. 1995,
    Cannizzo 1998, 2000),
   and may be more in line with either a disk-corona
  (Nayakshin et al. 2000)
 or disk-jet instability (McKinney \& Blandford 2009).
 Rea et al. (2011) derive stringent upper limits on the quiescent
X-ray emission from  Sw1955+26 using a $\sim$63 ks \emph{Chandra}
  observation, and use this to argue against a magnetar interpretation.
  \v{S}imon et al. (2012) present an analysis
of optical emission during the 2007 outburst
and find the optical emission manifests
as spikes which decrease in duration
as the burst progresses.
   They show that the emission can be explained
by pure synchrotron emission.
  The main part of the optical outburst following
the \emph{Swift}/BAT trigger lasted $\sim$0.3 d,
with subsequent ``echo outbursts'' between 1 and 3 d,
 as  in X-ray novae (Figure 8, from \v{S}imon et al. 2012). 
  The main part of the outburst
  has an exponential decay reminiscent of 
 the 1975 outburst in A0620-00,
but with a much shorter duration.
   If the global accretion disk
limit cycle (Cannizzo et al. 1995, 
       Dubus et al. 2001)
   is the mechanism
for this outburst, it suggests a much shorter
orbital period than the 7.75 hr for A0620-00 
 (McClintock \& Remillard 1986).

{\bf XRO 080109 - SN 2008D:}
  On 2008 January 9  \emph{Swift}/XRT  serendipitously discovered
 an
   extremely bright X-ray transient 
  (Soderberg et al. 2008)
   while undertaking a preplanned 
   observation of the galaxy NGC 2770 ($d=27$ Mpc).
 Two days earlier \emph{Swift}/XRT 
 had observed the same location
 and did not see a source.
X-ray outburst (XRO) 080109  lasted about 400 s
 and occurred in 
 one of the galaxy's spiral arms. 
  XRO 080109 was not a GRB (no $\gamma-$rays were detected),
and the total X-ray energy $E_X \simeq 2\times 10^{46}$ erg
was orders of magnitude lower than a GRB.
  The peak luminosity $\sim$$6\times10^{43}$ erg s$^{-1}$
is much greater than the Eddington luminosity
for a $\sim$$1\msun$ object, and also from type I X-ray
bursts. Therefore the standard accretion and   
  thermonuclear flash scenarios are excluded.

Simultaneous \emph{Swift}/UVOT 
 observations did not reveal a counterpart,
  but UVOT observations at 1.4 hr showed a brightening.
   Gemini North 8-m telescope 
 observations beginning at 1.7 d 
  revealed a spectrum   suggestive
of a young SN (Soderberg et al. 2008).
   Later observations
   confirmed the spectral features.
  The transient was classified as
a type Ibc SN based on the lack of H, and weak Si features.
 
\begin{figure}[h!]    
\begin{centering}
\includegraphics[height=3.45in]{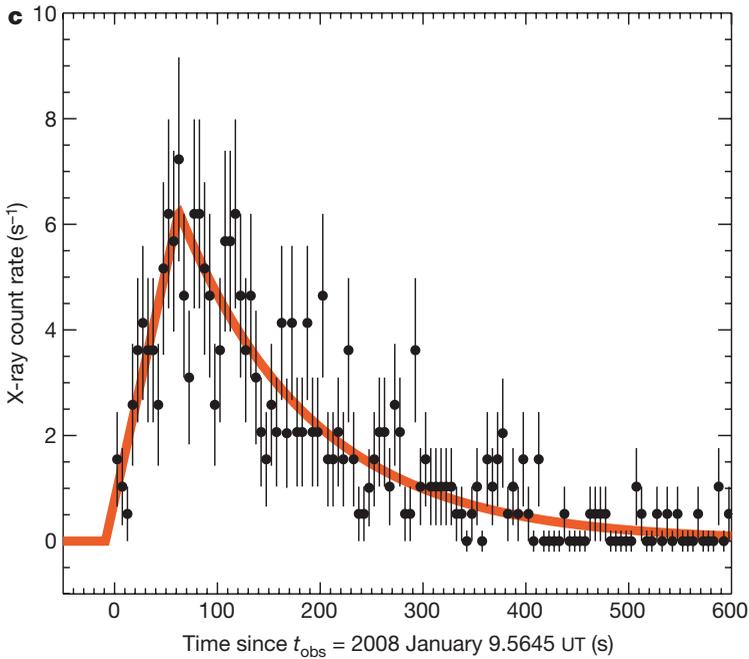}
\caption{X-ray light curve of XRO 080109/
SN 2008D (Soderberg et al. 2008).}
\end{centering}
 \end{figure}

  Soderberg et al. (2008)
  argue that the X-ray flash (Figure 9) indicates
 a trans-relativistic shock breakout
from a SN,
  where the radius at breakout
 is $\ga7\times 10^{11}$ cm,
 and the shock velocity at breakout is $\gamma\beta\la1.1$.
   Soderberg et al. (2008) estimate
a circumstellar density which yields an inferred pre-SN mass
loss rate $\sim10^{-5}\msunyr$,
  reinforcing the notion of a Wolf-Rayet progenitor.
 The similarity between the shock
break-out properties of the He-rich SN 2008D and the He-poor
GRB-associated SN 2006aj are consistent
 with  a dense stellar wind
around a compact Wolf-Rayet progenitor.

X-ray and radio observations 
presented  by  Soderberg et al. (2008)
  of SN 2008D are
the earliest ever obtained for
   a normal type Ibc SN.
  At
$t < 10$ d, the X-ray and peak radio 
 luminosities are orders
of magnitude less 
   than those of GRB afterglows
(Berger et al. 2003a,
   Frail et al. 2003),
   but comparable
 to those of normal type Ibc SN
  (Berger et al. 2003b,
  Kouveliotou et al. 2004).

{\bf EV Lac superflare:}
 On 25 Apr 2008 a hard X-ray superflare from the dMe star
 EV Lac triggered
 the BAT (Figure 10, from Osten et al. 2010).
  Flaring activity in this system had been seen previously
spanning radio to X-ray
wavelengths
using the VLA, \emph{HST}, and 
  \emph{Chandra} (Osten et al. 2005).
 The Apr 2008 event was the first large stellar flare from a 
  dMe flare star to cause a \emph{Swift}/BAT trigger
based on its hard X-ray intensity. 
  Its peak $0.3-100$ keV flux
of $5.3\times 10^{-8}$ erg cm$^{-2}$ s$^{-1}$
was $\sim$7000 times the star's quiescent
coronal flux (Osten et al. 2005, 2006).
  The soft X-ray spectrum
 of the flare shows evidence for 
  Fe K$\alpha$ emission at 6.4 keV.
 Osten et al. (2010) model
the K$\alpha$ emission
as fluorescence from the source of the flare
irradiating photospheric Fe, and derive
loop heights $h/R_*\simeq0.1$.
  It is interesting that
 with the sensitivity of BAT,
  one finds a small population
of  very energetic flares
producing hard X-ray flux at levels
commensurate with those seen from 
  GRBs.
  The frequency of flares this large
in M stars is unknown;
  a better understanding of the rate
would be an important factor
in determining the habitability
  of planets around M stars,
given the disastrous consequences
 of such a large energy release
 on the atmosphere of a nearby planet.

\begin{figure}[h!]    
\begin{centering}
\includegraphics[height=3.45in]{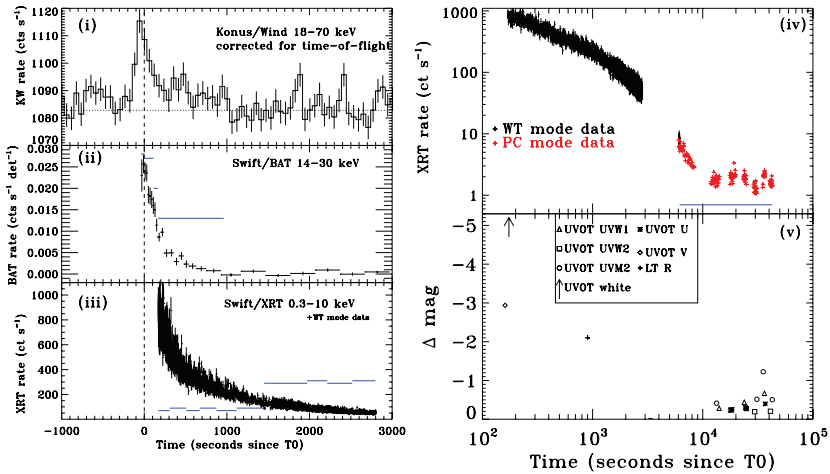}
\caption{Light curves during the flare from EV Lac
  (Osten et al. 2010).} 
\end{centering}
 \end{figure}

{\bf Pulsing GRB 090709:}
 Analysis of
the \emph{Swift}/BAT data
  revealed a quasiperiodic signal 
at  8.06 s, with a $Q-$factor of $\sim$11 
       (Markwardt et al. 2009).
Markwardt et al. discuss a magnetar
scenario, with 8 s representing
  the pulsar spin. 
 G\"otz et al. (2009)
   confirmed the periodicity
  with a finding of a 8.11 s signal in
 \emph{INTEGRAL}/SPI-ACS data.
  However,  detailed follow-up work suggests
a more pedestrian scenario $-$ a 
 standard long GRB 
 (Figure 11, from de Luca et al. 2010;
    Cenko et al. 2010).
 De Luca et al. (2010)
   reanalyzed the  \emph{Swift} 
and  \emph{INTEGRAL} data and excluded
  any significant modulation
 at 8.1 s. Their fitting of 
  \emph{Swift}/XRT 
    and \emph{XMM-Newton}/EPIC
  X-ray spectra imply a redshift 
 $z\sim$$4-5$, too far to be a magnetar.
  They also note the lack of short ($\la 0.5$s) and
hard very bright initial spike that are seen
in SGR giant flares, and the lack of an obvious nearby galaxy
progenitor. The huge energy requirement implied by the
apparent cosmological distance works against the SGR giant
  flare hypothesis.
  Cenko et al. (2010) present broadband  observations
 of GRB 090709 and also conclude it was
probably a standard long GRB at cosmological
distances. They detect the periodic signal
reported by Markwardt et al. (2009) and G\"otz et al.
 (2009) at only $\sim$$2\sigma$ significance.
  Perley et al. (2010)
 discovered a faint galaxy at the afterglow location
 ($K' = 22.0 \pm 0.2$ mag), 
 confirming
the extragalactic, cosmological nature of the burst.
   To date, a firm spectroscopic redshift 
has not been obtained.

\begin{figure}[h!]
\begin{centering}
\includegraphics[height=3.45in]{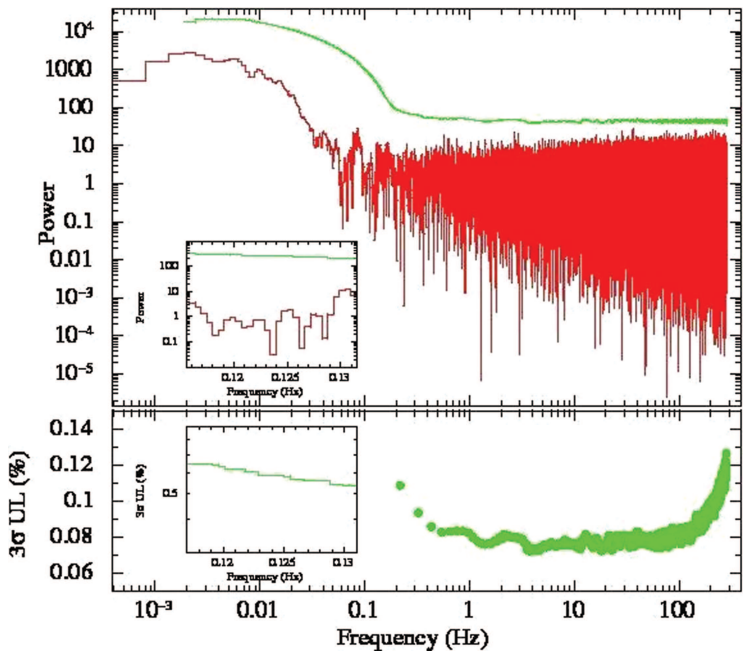}
\caption{The upper panel indicates 
  the power spectrum for 
 the \emph{Swift}/XRT light curve
  of GRB 090709  between  $T0 + 77$
  and $T0 + 539$ s,
   along
with the threshold for the detection of 
  sinusoidal signals at the $3\sigma$
confidence level.
  The lower panel shows the upper limits 
 on the pulsed fraction. 
  The XRT data are in 
  WT mode
  (from De Luca et al. 2010).}
\end{centering}
 \end{figure}

{\bf GRB 101225 ``Christmas burst'':}
 GRB 101225 was quite unusual:
 it had a $T_{90} > 1700$ s and
exhibited a curving decay when plotted
  in the traditional $\log F - \log t$ 
 coordinates.
The total BAT fluence was $\ga3\times 10^{-6}$
   erg cm$^{-2}$.
  The  XRT and UVOT 
 found a bright, long-lasting counterpart.
  Ground based telescopes followed
the event, mainly in $R$ and $I$, 
  and failed to detect any spectral features.
 At later times a color change from
 blue to red was seen;
  \emph{HST} observations at 20 d
  found a very red object
 with no apparent host.
Observations from the Spanish 
  Gran Telescopio Canarias
at 180 d
 detected GRB 101225  at 
  $g_{\rm AB}= 27.21\pm0.27$ mag 
  and
  $r_{\rm AB}= 26.90\pm0.14$ (Th\"one et al. 2011). 
  Considered together, 
   these characteristics  
are unique to this burst (Figure 12), and
led Campana et al. (2011)
 to propose that it was caused by  a minor body
  like an asteroid or comet
  becoming disrupted
and accreted by a NS.
    Depending on its composition,
 the tidal disruption
  radius would be $\sim10^5 - 10^6$ km.
  Campana et al. find an adequate fit to the
light curve by positing a $\sim 5\times 10^{20}$ g
asteroid
   with a periastron radius $\sim9000$ km. 
  If half the asteroid mass is accreted
 they derive a total fluence $4\times 10^{-5}$
 erg cm$^{-2}$
  and distance $\sim3$ kpc. 

\begin{figure}[h!]
\begin{centering}
\includegraphics[height=3.45in]{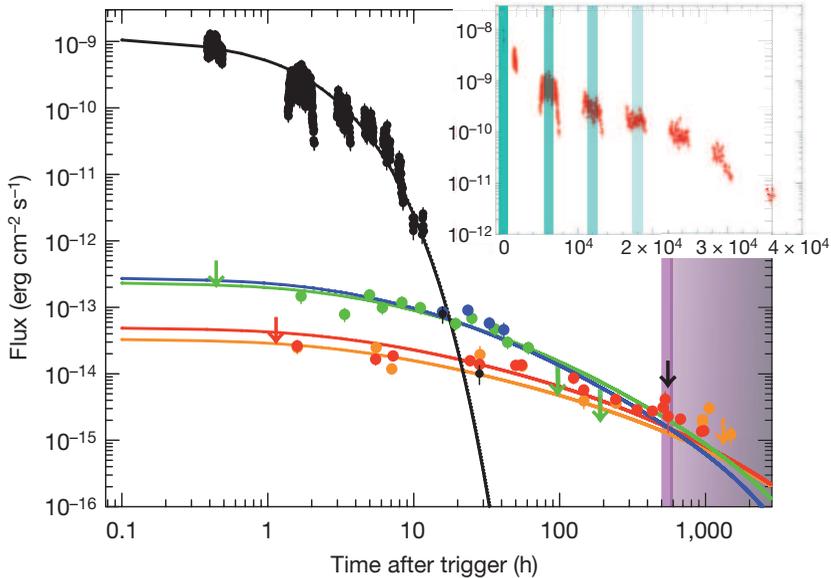}
\caption{Light curves of GRB 101225
 in  five
energy bands: X-rays at 1 keV (black), 
  UV at 2030 \AA \ (green) 
   and 2634 \AA \ 
(blue), and 
  optical at 6400 \AA \ ($R$ band, red) 
  and 7700 \AA  \
    ($I$ band, orange)
 (Campana et al. 2011).
 The inset (with $t$ in seconds)
  shows the \emph{Swift}/XRT
  light curve. 
 Shaded regions highlight the periastron
passages calculated using a tidal disruption model
 (Campana et al. 2011).
}
\end{centering}
 \end{figure}

Th\"one et al. (2011)
  offer   a different explanation for  GRB 101225 $-$
  the  merger of a He star and a NS
   leading to a concomitant SN.
They derive a pseudo-redshift
  $z=0.33$ by fitting the spectral-energy
distribution and light curve of the optical emission with a GRB-
 SN template.
  Thus in their interpretation the event was much more distant
and energetic.
  They argue for the presence of a faint, unresolved
  galaxy in deep optical observations,
   and fit the long term light curve
  with a template of the broad-line
  type Ic SN 1998bw associated with GRB 980425.
  If their distance is correct as well as their interpretation
 of a component emerging at 10 d as being a SN, 
 then its 
absolute peak magnitude $M_{\rm V, \ abs} = -16.7$ mag
  would  make it
the faintest SN associated with a long GRB.
  The isotropic-equivalent energy release at $z=0.33$ would be
  $>1.4\times 10^{51}$ erg
  which is typical of other long GRBs but
  greater than most other
low-redshift GRBs associated with SNe.

\section{Summary}

The last fifty years has been an exciting time of
great discoveries in high energy astrophysics,
  enabled by innovative advances in detectors.
 The universe has been revealed to be a more
  wondrous and sometimes violent
  place than previously imagined,
  from $\gamma-$ray bursts on the most distant scales,
  to 
  terrestrial $\gamma-$ray flashes
   right here on Earth.
    We have been surprised by the 
   seemingly low energy phenomena that turned out
 to have high energy emission associated with them,
  as well as the discoveries of entirely new phenomena.
  Given our  incomplete  understanding
 of many of these phenomena, there is enormous
opportunity for more detailed follow-up.
 The future is bright for time domain astrophysics.


\begin{thebibliography}


\bibitem{}
Atwood, W.~B. et al. 2009 
The Large Area Telescope on 
the Fermi Gamma-Ray Space Telescope Mission.
 \textit{Ap. J.} \textbf{697}, 1071--1102.

\bibitem{}
Berger, E., Kulkarni, S.~R. \& Frail, D.~A. 2003a
A Standard Kinetic Energy Reservoir in Gamma-ray Burst Afterglows.
 \textit{Ap. J.} \textbf{590}, 379--385.

\bibitem{}
Berger, E., Kulkarni, S.~R., Frail, D.~A.
  \& Soderberg, A.~M.  2003b
A Radio Survey of Type Ib and Ic Supernovae:
   Searching for Engine-driven Supernovae.
 \textit{Ap. J.} \textbf{599}, 408--418.

\bibitem{} 
Bloom, J.~S. et al. 2011
A Possible Relativistic Jetted Outburst
from a Massive Black Hole Fed by a
Tidally Disrupted Star.
 \textit{Science} \textbf{333}, 203--205.

\bibitem{} 
        Bloom, J.~S., 
   Djorgovski, S.~G.
  \& Kulkarni, S.~R. 2001
The Redshift and the Ordinary Host Galaxy of GRB 970228.
 \textit{Ap. J.} \textbf{554}, 678--683.

\bibitem{}
   Burrows, D.~N. et al. 
 2006
   Jet Breaks in Short Gamma-Ray Bursts. 
  II. The Collimated Afterglow of GRB 051221A.
\textit{Ap. J.} \textbf{653}, 468--473.

\bibitem{} 
   Burrows, D.~N. et al. 2011
Relativistic jet activity from the tidal disruption of
a star by a massive black hole.
 \textit{Nature} \textbf{476}, 421--424.

\bibitem{}
Campana, S. et al. 2006 
The association of GRB 060218 with a supernova 
 and the evolution of the shock wave.
 \textit{Nature} \textbf{476}, 421--424.

\bibitem{}
Campana, S. et al. 2011 
The unusual gamma-ray burst GRB 101225A 
  explained as a minor body falling onto a neutron star.
 \textit{Nature} \textbf{480}, 69--71.

\bibitem{}
Cannizzo, J.~K. 1998
The Accretion Disk Limit Cycle Mechanism in the 
Black Hole X-Ray Binaries: Toward an Understanding 
of the Systematic Effects.
\textit{Ap. J.} \textbf{494}, 366--380.

\bibitem{}
Cannizzo, J.~K. 2000
On the Role of Irradiation and Evaporation in 
Strongly Irradiated Accretion Disks in the 
Black Hole X-Ray Binaries: Toward an Understanding 
of FREDs and Secondary Maxima.
\textit{Ap. J.} \textbf{534}, L35--L38.

\bibitem{}
Cannizzo, J.~K., Chen, W. \& Livio, M. 1995
The Accretion Disk Limit Cycle Instability 
 in Black Hole X-Ray Binaries.
\textit{Ap. J.} \textbf{454}, 880--894.

\bibitem{}
Cannizzo, J.~K., Lee, H.~M. \& Goodman, J 1990
 The Disk Accretion of a Tidally Disrupted Star
  onto a Massive Black Hole.
\textit{Ap. J.} \textbf{351}, 38--46.

\bibitem{}
Cannizzo, J.~K., Troja, E. \& Lodato, G. 2011
GRB 110328A/Swift J164449.3+573451: 
The Tidal Obliteration of a Deeply Plunging Star?
\textit{Ap. J.} \textbf{742}, 32--38.

\bibitem{}
Cenko, S.~B. et al. 2008
GRB 070125: The First Long-Duration Gamma-Ray Burst
  in a Halo Environment.
\textit{Ap. J.} \textbf{677}, 441--447.

\bibitem{}
Cenko, S.~B.  et al. 2010 
Unveiling the Origin of GRB 090709A: Lack of
Periodicity in a Reddened Cosmological
Long-Duration Gamma-Ray Burst.
 \textit{A. J.} \textbf{140}, 224--234.

\bibitem{}
Chandra, P. et al. 2008
  Comprehensive Study of GRB 070125, 
 A Most Energetic Gamma-Ray Burst.
\textit{Ap. J.} \textbf{683}, 924--942.

\bibitem{} 
   Chupp, E.~L.,
 Forrest, D.~J., 
  Higbie, P.~R.,
    Suri, A.~N.,
    Tsai, C. 
 \& Dunphy, P.~P. 1973 
 Solar gamma Ray Lines observed during 
 the Solar Activity of August 2 to August 11, 1972.
 \textit{Nature} \textbf{241}, 333--335.

\bibitem{bib2}
Cline, T.~L.  et al. 1980
  Detection of a Fast, Intense and
   Unusual Gamma-ray Transient.
 \textit{Ap. J.} \textbf{237}, L1--L5.

\bibitem{}
 della Valle, M.  et al. 2006
 An enigmatic long-lasting $\gamma-$ray burst not accompanied
               by a bright supernova.
 \textit{Nature} \textbf{444}, 1050--1052.

\bibitem{}
  de Luca, A.,
    Esposito, P.,
    Israel, G.~L.,
   G\"otz, D.,
    Novara, G., 
     Tiengo, A. 
 \& Mereghetti, S.
 2010
 \emph{XMM-Newton} and \emph{Swift}
   observations prove 
 GRB090709A to be a distant, 
                        standard, long GRB.
 \textit{M. N. R. A. S.} \textbf{402}, 1870--1876.


\bibitem{}
   Dotani, T., 
   Hayashida, K., 
  Inoue, H.,
  Itoh, M. \& 
  Koyama, K.
  1987
  Discovery of an unusual hard X-ray source in the region of supernova 1987A.
 \textit{Nature} \textbf{330}, 230--231.



\bibitem{}
Dubus, G., Hameury, J.-M. \& Lasota, J.-P.
  2001
 The disc instability model for X-ray transients: Evidence
         for truncation and irradiation.
  \textit{A. \& A.} \textbf{373}, 251--271.

\bibitem{}
 Dwyer, J.~R. 
 2003 
 A fundamental limit on electric fields in air.
\textit{Geophysical Research Letters} \textbf{30}, 2055--2058. 

\bibitem{}
Fong, W.-F. et al. 
 2012 
  A Jet Break in the X-ray Light Curve 
  of Short GRB 111020A: 
Implications for Energetics and Rates.
   \textit{astro-ph} 1204.5475.

\bibitem{}
 Frail, D.~A., Kulkarni, S.~R., Berger, E.
   \& Wieringa, M.~H. 2003
A Complete Catalog of Radio Afterglows:
    The First Five Years.
\textit{A. J.} \textbf{125}, 2299--2306.

\bibitem{}
 Fynbo, J.~P.~U. et al. 2006
 No supernovae associated with two long-duration
   $\gamma-$ray bursts.
\textit{Nature} \textbf{444}, 1047--1049.

\bibitem{}
 Gallagher, J.~S. \& Starrfield, S.
 1978
 Theory and Observations of Classical Novae.
\textit{A. R. A. \& A.} \textbf{16}, 171--214.

\bibitem{}
  Gal-Yam, A. et al. 2006
A novel explosive process is required for the
  $\gamma-$ray burst GRB060614.
 \textit{Nature} \textbf{444}, 1053--1055.

\bibitem{}
Gehrels, N. et al. 2004  
The Swift Gamma-Ray Burst Mission.
\textit{Ap. J.} \textbf{611}, 1005--1020.

\bibitem{}
Gehrels, N. et al. 2006  
 A new $\gamma-$ray burst classification scheme
                 from GRB 060614.
\textit{Nature} \textbf{444}, 1044--1046.

\bibitem{}
Gehrels, N. et al. 2008
Correlations of Prompt and Afterglow Emission 
 in Swift Long and Short Gamma-Ray Bursts.
\textit{Ap. J.} \textbf{689}, 1161--1172.

\bibitem{}
Gehrels, N., Ramirez-Ruiz, E. \& Fox, D.~B. 2009
  Gamma-Ray Bursts in the \emph{Swift} Era.
\textit{A. R. A. \& A.} \textbf{47}, 567--617.

\bibitem{}
Gezari, S. et al. 2006
  Ultraviolet Detection of the Tidal Disruption of a Star
  by a Massive Black Hole.  
  \textit{Ap. J.} \textbf{653}, L25--L28.

\bibitem{}
Gezari, S. et al. 2008
UV/Optical Detections of Candidate 
 Tidal Disruption Events by GALEX and CFHTLS1.
  \textit{Ap. J.} \textbf{676}, 944--969.

\bibitem{}
 G\"otz, D.,
   Mereghetti, S.,
  von Kienlin, A. 
      \& Beck, M.  2009
 GRB 090709A: confirmation of the 
  periodicity in the SPI-ACS data.
 \textit{GCN}, 9649.

\bibitem{}
    Grupe,            D.,
  Burrows,         D.~N.,
  Patel,           S.~K.,
   Kouveliotou,       C., 
   Zhang,             B., 
  M{\'e}sz{\'a}ros,   P.,
   Wijers,      R.~A.~M.,
    \& Gehrels, N.
2006
    Jet Breaks in Short Gamma-Ray Bursts. 
 I. The Uncollimated Afterglow of GRB 050724.
   pages = {462-467},
  \textit{Ap. J.} \textbf{653}, 462--467.

\bibitem{}
Gurevich, A.~V.,  Milikh, G.~M.  \&  Roussel-Dupr\'e, R. 1992 
 Runaway electron mechanism of air breakdown and preconditioning 
                     during a thunderstorm.
\textit{Phys. Lett. A} \textbf{165}, 463--468.

\bibitem{}
  Holland, S.~T. et al.
    2005
 GRB050925: Swift-BAT detection of a short burst.
  periodicity in the SPI-ACS data.
 \textit{GCN}, 4034.

\bibitem{}
   in't Zand, J.~J.~M.,  
    Kuulkers, E.,        
     Bazzano, A.,        
  Cornelisse, R.,        
      Cocchi, M.,        
       Heise, J.,        
      Muller, J.~M.,     
    Natalucci, L.,       
    Smith, M.~J.~S.  \&  
      Ubertini, P.      
     2000
\emph{BeppoSAX} observations of the nearby 
 low-mass X-ray binary and fast 
  transient SAX J1819.3-2525.
\textit{A. \& A.} \textbf{357}, 520--526.

\bibitem{}
 Kasliwal, M.~M.  et al. 2008,
   GRB 070610: A Curious Galactic Transient.
  \textit{Ap. J.} \textbf{678}, 1127--1135.

\bibitem{} 
Klebesadel, R.~W., Strong, I.~B.  \& Olson, R.~A. 1973
  Observations of Gamma-Ray Bursts of Cosmic Origin.
 \textit{Ap. J.} \textbf{182}, L85--L88.

\bibitem{}
Komossa, S., Halpern, J.,
Schartel, N.; Hasinger, G., 
Santos-Lleo, M. \& Predehl, P.
  2004
   A Huge Drop in the X-Ray Luminosity of the 
 Nonactive Galaxy RX J1242.6-1119A, and the First 
    Postflare Spectrum: 
 Testing the Tidal Disruption Scenario.
     \textit{Ap. J.} \textbf{603}, L17--L20.


\bibitem{} 
Kouveliotou, C., 
   Meegan, C.~A., Fishman, G.~J., Bhat, N.~P.,
   Briggs, M.~S.,      Koshut, T.~M., 
 Paciesas, W.~S. \& Pendleton, G.~N. 
 1993
Identification of two classes of gamma-ray bursts.
 \textit{Ap. J.} \textbf{413}, L101--L104.

\bibitem{}
Kouveliotou, C., Strohmayer, T., 
 Hurley, K., van Paradijs, J.,
 Finger, M.~H., Dieters, S., 
 Woods, P., Thompson, C. \& Duncan, R.~C.
 1999
Discovery of a Magnetar Associated with the 
 Soft Gamma Repeater SGR 1900+14.
 \textit{Ap. J.} \textbf{510}, L115--L118.

\bibitem{}
Kouveliotou, C. et al.
 2004
 \emph{Chandra} Observations of the X-ray Environs
    of SN 1998bw/GRB 980425.
 \textit{Ap. J.} \textbf{608}, 872--882.

\bibitem{}
Levan, A.~J. et al. 
 2011
 An Extremely Luminous Panchromatic Outburst 
 from the Nucleus of
a Distant Galaxy.
 \textit{Science} \textbf{333}, 199--202.

\bibitem{}
  Markwardt,  C.~B.
   et al.
 2005
 Refined analysis of the Swift-BAT soft short burst.
 \textit{GCN}, 4037.


\bibitem{}
     Markwardt,   C.~B., 
     Gavriil,     F.~P.,
     Palmer,      D.~M.,
     Baumgartner, W.~H.
  \& Barthelmy,   S.~D. 2009
  GRB 090709A: Quasiperiodic 
 variations in the BAT light curve.
 \textit{GCN}, 9645.

\bibitem{} 
    Matz,  S.~M.,
    Share, G.~H., 
  Leising, M.~D.,
    Chupp, E.~L. \&
  Vestrand, W.~T.	
  1988
  Gamma-ray line emission from SN1987A.
 \textit{Nature} \textbf{331}, 416--418.


\bibitem{} 
Mazets, E.~P., Golentskii, S.~V., 
   Ilinskii, V.~N., Aptekar, R.~L. \&
     Guryan, Iu.~A.  1979 
  Observations of a flaring X-ray pulsar in Dorado.
 \textit{Nature} \textbf{282}, 587--589.

\bibitem{}
 Mazzali, P.~A.,
  Deng, J.,
  Nomoto, K.,
  Sauer, D.~N.,
 Pian, E.,
 Tominaga, N.,
 Tanaka, M.,
  Maeda, K. \& 
 Filippenko, A.~V. 
  2006
  A neutron-star-driven X-ray flash associated with supernova SN 2006aj.
\textit{Nature} \textbf{442}, 1018--1020.

\bibitem{}
  McClintock, J.~E. \& Remillard, R.~A. 1986
  The Black Hole Binary A0620-00.
 \textit{Ap. J.} \textbf{308}, 110--122.

\bibitem{}
   McKinney, J.~C. \& Blandford, R.~D. 2009
  Stability of relativistic jets from rotating, accreting
   black holes via fully three-dimensional magnetohydrodynamics
    simulations.
 \textit{M. N. R. A. S.} \textbf{394}, L126--L130.

\bibitem{}
  Nayakshin, S.,  Rappaport, S.
  \& Melia, F. 2000
   Time-dependent Disk Models for the Microquasar GRS 1915+105.
  \textit{Ap. J.} \textbf{535}, 833--852.

\bibitem{}
  Norris, J.~P.  \& Bonnell, J.~T.
 2006
 \textit{Ap. J.} \textbf{643}, 266--275.

\bibitem{}
Onda, K. et al. 2010 
Time-Resolved Spectral Variability of the Prompt 
Emission from GRB 070125
Observed with Suzaku/WAM.
 \textit{P. A. S. J.} \textbf{62}, 547--556.

\bibitem{}
Orosz, J.~ A.,        
 Kuulkers, E.,        
   van der Klis, M.,  
 McClintock, J.~E.,   
 Garcia, M.~R.,       
  Callanan, P.~J.,    
   Bailyn, C.~D.,     
   Jain, R.~K.        
 \& Remillard, R.~A.  
 2001
 A Black Hole in the Superluminal 
 Source SAX J1819.3-2525 (V4641 Sgr).
 \textit{Ap. J.} \textbf{555} 489--503.

\bibitem{}
  Osborne, J.~P. et al.
  2011
  The Supersoft X-ray Phase of Nova RS Ophiuchi 2006.
 \textit{Ap. J.} \textbf{727}, 124--133.

\bibitem{}
Osten, R.~A.,
Hawley, S.~L.,
Allred, J.,
Johns-Krull, C.~M.,
Brown, A. \&
Harper, G.~M.  
  2006
From Radio to X-Ray: The Quiescent Atmosphere 
of the dMe Flare Star EV Lacertae.
 \textit{Ap. J.} \textbf{647}, 1349--1374.

\bibitem{}
Osten, R.~A.,
Hawley, S.~L.,
Allred, J.~C.,
Johns-Krull, C.~M.
  \& Roark, C.
  2005
From Radio to X-Ray: Flares on
the dMe Flare Star EV Lacertae.
 \textit{Ap. J.} \textbf{621}, 398--416.

\bibitem{}
Osten, R.~A. et al. 2010
The Mouse That Roared: A Superflare from the 
dMe Flare Star EV Lac Detected by \emph{Swift} and \emph{Konus-Wind}.
 \textit{Ap. J.} \textbf{721}, 785--800.

\bibitem{bib3} 
Palmer, D.~M. et al. 2005
  A giant $\gamma-$ray flare from the magnetar SGR 1806–20.
 \textit{Nature} \textbf{434}, 1107--1109.

\bibitem{}
    Perley, D.~A.,
    Cenko, S.~B. 
 \& Bloom, J.~S.
 2010 
  GRB 090709A: Host galaxy detection.
 \textit{GCN},  10903.

\bibitem{}
 Phinney, E.~S. 	
 1989
 Manifestations of a Massive Black Hole in the Galactic Center.
 \textit{IAU Symposium $-$ The Center of the Galaxy}
 ed. M. Morris, \textbf{136}, 543--553.

\bibitem{}
   Pian, E. et al.
 2006
An optical supernova associated with the
              X-ray flash XRF 060218.
 \textit{Nature} \textbf{442}, 1011--1013.



\bibitem{}
   Pinto, P.~A. \& 
  Woosley, S.~E.
1988a
  The theory of gamma-ray emergence in supernova 1987A.
 \textit{Nature} \textbf{333}, 534--537.

\bibitem{}
   Pinto, P.~A. \& 
  Woosley, S.~E.
1988b
  X-ray and gamma-ray emission from supernova 1987A.
 \textit{Ap. J.} \textbf{329}, 820--830.



\bibitem{}
  Rea, N.,
  Jonker, P.~G.,
  Nelemans, G.,
  Pons, J.~A., 
  Kasliwal, M.~M. 
  Kulkarni, S.~R. \&
  Wijnands, R.	
 2011
  The X-ray Quiescence of Swift J195509.6+261406
 (GRB 070610): An Optical Bursting X-ray Binary?
\textit{Ap. J.} \textbf{729}, L21--L25.	

\bibitem{}
  Remillard, R.~A.  \& McClintock, J.~E.
 2006 
  X-Ray Properties of Black-Hole Binaries.
 \textit{A. R. A. \& A.} \textbf{44}, 49--92.

\bibitem{} 
Rees, M.~J. 1988
 Tidal disruption of stars by black holes of
 $10^6-10^8$ solar masses in nearby galaxies.
 \textit{Nature} \textbf{333}, 523--528.

\bibitem{}
   Rothschild, R.~E., 
 Lingenfelter, R.~E.,
       Seward, F.~D. 
  \&  Vancura, O.
  1993
 An X-ray counterpart to the 
 5 March 1979 gamma ray burst?
 \textit{The Compton Gamma Ray Observatory; 
   American Institute of Physics}
  ed. M. Friedlander, N. Gehrels \&
    R.~J. Macomb, \textbf{280}, 808--812.

\bibitem{}
  Rothschild, R.~E., Kulkarni, S.~R. 
 \& Lingenfelter, R.~E.
1994
   Discovery of an X-ray source coincident 
 with the soft $\gamma-$ray repeater 0525-66.
 \textit{Nature}  \textbf{368}, 432--434.

\bibitem{} 
Shakura, N.~I. \& Sunyaev, R.~A. 1973
Black Holes in Binary Systems.
          Observational Appearance.
\textit{A. \& A.} \textbf{24}, 337--355.

\bibitem{}
  \v{S}imon, V. et al.
2012    
Outburst and Flares from the unique source SWIFT J1955+2614.
\textit{M. N. R. A. S.} \textbf{422}, 981--989.

\bibitem{}
Smith, D.~M., Lopez, L.~I., Lin, R.~P. 
 \& Barrington-Leigh, C.~P. 2005 
Terrestrial gamma-ray flashes observed up to 20 MeV.
\textit{Science} \textbf{307}, 1085--1088.

\bibitem{} Soderberg, A.~M. et al.
 2006
 Relativistic ejecta from X-ray flash XRF 060218 
and the rate of cosmic explosions.
 \emph{Nature} \textbf{442}, 1014--1017.

\bibitem{} Soderberg, A.~M. et al.
 2008
 An extremely luminous X-ray outburst at the birth of
  a supernova.
 \emph{Nature} \textbf{453}, 469--474.


\bibitem{} 
   Sunyaev, R. et al. 
  1987
    Discovery of hard X-ray emission from supernova 1987A.
 \emph{Nature} \textbf{330}, 227--229.


\bibitem{}
  Tanaka, Y.  \& Shibazaki, N.
 1996
 X-ray Novae.
\textit{A. R. A. \& A.} \textbf{34}, 607--644.

\bibitem{} 
 Teegarden, B.~J., 
 Barthelmy, S.~D., 
 Gehrels, N., Tueller, J. \& Leventhal, M. 1989
Resolution of the 1,238-keV gamma-ray line from supernova 1987A.
\textit{Nature} \textbf{339}, 122--123.

\bibitem{}
Th\"one, C.~C. et al. 2011 
The unusual $\gamma-$ray burst GRB 101225A from a helium
star/neutron star merger at redshift 0.33.
 \textit{Nature} \textbf{480}, 72--74.

\bibitem{}
Thompson, C. \& Duncan, R.~C. 1995
  The Soft Gamma Repeaters as Very Strongly Magnetized
   Neutron Stars $-$ I. Radiative  Mechanism for Outbursts.
\textit{M. N. R. A. S.} \textbf{275}, 255--300.

\bibitem{}
Thompson, C. \& Duncan, R.~C. 2001
  The Giant Flare of 1998 August 27 from SGR 1900+14. II. Radiative
  Mechanism and Physical Constraints on the Source.
\textit{Ap. J.} \textbf{561}, 980--1005.

\bibitem{} 
van Paradijs, J. et al. 1997
Transient optical emission from the error box 
of the $\gamma-$ray burst of 28 February 1997.
\textit{Nature} \textbf{386}, 686--689. 

\bibitem{}
Winkler, C. et al. 2003
The INTEGRAL mission.
 \textit{A. \& A.} \textbf{411}, L1--L6.


\end{thebibliography}
\end{document}